%% file: main.tex
\documentclass[conference]{IEEEtran}
\usepackage{cite}
\usepackage{amsmath,amssymb,amsfonts}
\usepackage{algorithmic}
\usepackage{graphicx}
\usepackage{textcomp}
\usepackage{url}
\usepackage{enumitem}
\usepackage{xcolor}
\def\BibTeX{{\rm B\kern-.05em{\sc i\kern-.025em b}\kern-.08em
    T\kern-.1667em\lower.7ex\hbox{E}\kern-.125emX}}
\usepackage{xspace}
\usepackage{comment}
\usepackage{stfloats}
\usepackage{tabularx}
\usepackage{booktabs}
\usepackage{caption}
\usepackage{xurl}
\usepackage[hidelinks]{hyperref}

\begin{document}

\newcommand{\rasa}{\textsc{RASA}\xspace}
\newcommand{\dataset}{\textsc{BRASATO}\xspace}
\newcommand{\datasetFullName}{\textsc{Bot RASA collecTiOn}\xspace}

\newcommand{\bigDataset}{\textsc{TOFU-R}\xspace}

\newcommand{\bigDatasetFullName}{\textsc{Rasa Task-based chatbOts From githUb}\xspace}

\newcommand{\initialNumber}{\color{black}\textsc{8,634}\xspace\color{black}}
\newcommand{\GRCSNumber}{\color{black}\textsc{5,271}\xspace\color{black}}
\newcommand{\BRASATONumber}{\color{black}\textsc{193}\xspace\color{black}}
\newcommand{\urlBigDataset}{\href{https://gitlab.com/securitychatbot/rasa-chatbot-dataset}{\url{https://gitlab.com/securitychatbot/rasa-chatbot-dataset}}}

\newcommand{\urlDataset}{\href{https://gitlab.com/securitychatbot/rasa-chatbot-dataset}{\url{https://gitlab.com/securitychatbot/rasa-chatbot-dataset}}}

\makeatletter
\newcommand{\linebreakand}{%
  \end{@IEEEauthorhalign}
  \hfill\mbox{}\par
  \mbox{}\hfill\begin{@IEEEauthorhalign}
}
\makeatother

\title{
Towards the Assessment of Task-based Chatbots: From the TOFU-R Snapshot to the BRASATO Curated Dataset\\
}

\author{\IEEEauthorblockN{Elena Masserini}
\IEEEauthorblockA{\textit{University of Milano-Bicocca} \\
Milan, Italy \\
elena.masserini@unimib.it}
\and
\IEEEauthorblockN{Diego Clerissi}
\IEEEauthorblockA{\textit{University of Milano-Bicocca} \\
Milan, Italy \\
diego.clerissi@unimib.it}
\and
\IEEEauthorblockN{Daniela Micucci}
\IEEEauthorblockA{\textit{University of Milano-Bicocca} \\
Milan, Italy \\
daniela.micucci@unimib.it}
\linebreakand
\IEEEauthorblockN{João R. Campos}
\IEEEauthorblockA{\textit{University of Coimbra, CISUC/LASI, DEI} \\
Coimbra, Portugal \\
jrcampos@dei.uc.pt}
\and
\IEEEauthorblockN{Leonardo Mariani}
\IEEEauthorblockA{\textit{University of Milano-Bicocca} \\
Milan, Italy \\
leonardo.mariani@unimib.it}

}

\maketitle

\begin{abstract}
Task-based chatbots are increasingly being used to deliver real services, yet assessing their reliability, security, and robustness remains underexplored, also due to the lack of large-scale, high-quality datasets. 
The emerging automated quality assessment techniques targeting chatbots often rely on limited pools of subjects, such as custom-made toy examples, or outdated, no longer available, or scarcely popular agents, complicating the evaluation of such techniques. 
In this paper, we present two datasets and the tool support necessary to create and maintain these datasets.
The first dataset is
\bigDatasetFullName (\bigDataset), which is a snapshot of the Rasa chatbots available on GitHub, representing the state of the practice in open-source chatbot development with Rasa. The second dataset is \datasetFullName (\dataset), a curated selection of the most relevant chatbots for dialogue complexity, functional complexity, and utility, 
whose goal is to ease reproducibility and facilitate research on chatbot reliability.
\end{abstract}



\section{Introduction} \label{sec:introduction}
\input{introduction} 

\section{Rasa Chatbots} \label{sec:rasa}
\input{rasa} 

\section{Methodology} \label{sec:methodology}
\input{methodology}

\section{Descriptive Analysis of the Two Datasets} \label{sec:dataset}
\input{dataset}

\section{Related Work} \label{sec:related}
\input{related}

\section{Conclusions} \label{sec:conclusions}
\input{conclusions}

\smallskip 

\emph{Acknowledgment}. This work has been partially supported by the Engineered MachinE Learning-intensive IoT systems (EMELIOT) national research project,
which has been funded by the MUR under the PRIN 2020 program
(Contract nr. 2020W3A5FY), and through national funds by FCT I.P., in the framework of the Project UIDB/00326/2025 and UIDP/00326/2025.

\bibliographystyle{IEEEtran}
\bibliography{IEEEabrv,references}

\end{document}

%% file: introduction.tex
Chatbot technology is rapidly evolving~\cite{sanchez2024automating,adamopoulou2020chatbots}. Services are becoming more and more available as chatbots that can understand natural language and provide 
support in response to textual or voice requests made by users. Notable examples are help desk chatbots~\cite{fiore2019forgot}, food delivery chatbots~\cite{de2021s}, and tourism assistance chatbots~\cite{ukpabi2019chatbot}. 
In contrast to general-purpose chatbots (e.g., ChatGPT\footnote{\url{https://chatgpt.com/}} and Gemini\footnote{\url{https://gemini.google.com/}}), chatbots designed to deliver actual services are referred to as \emph{Task-based Chatbots}~\cite{grudin2019chatbots,adamopoulou2020chatbots}. 

Task-based chatbots implement a \emph{Natural-Language Understanding Pipeline} that can interpret the users' messages, detecting the user's intent, and extracting the relevant content from sentences (e.g., the intended date for a reservation). Some of the operations implemented in the pipeline are performed by exploiting \emph{machine learning models}, either pre-trained or trained on a specific set of sentences. Users' messages may activate \emph{actions}, which implement the business logic controlled by the chatbot agent, often depending on external APIs or libraries.
Task-based agents can be implemented using different platforms. For instance, Google DialogFlow~\cite{dialogflow} and Amazon Lex~\cite{amazon-lex} are popular commercial platforms, while Rasa~\cite{rasa} is a popular open-source platform~\cite{abdellatif2021comparison,cao2023characterizing}.

Despite their increasing popularity, ensuring the reliability of task-based agents is a largely open challenge
~\cite{lambiase2024motivations,cabot2021testing}. For instance, testing agents requires generating natural language sentences that can activate the right dialogues and oracles that can precisely interpret the (natural language) responses. 
Assessing the security of task-based agents requires exploring the capability of the agents to interpret requests without being misled by the natural language, as well as designing techniques that can analyze the implementation of task-based agents to reveal potential vulnerabilities. Similarly, assessing the robustness of task-based agents requires new models and approaches that consider the conversation as a first-class entity.

So far, contributions have been limited in scope and depth~\cite{cabot2021testing,lambiase2024motivations,li2022review}. Some approaches investigated white-box generation of test cases\cite{bravo2020testing,canizares2024coverage,rapisarda2025test}, 
while others focused on black-box robustness testing, introducing syntactic and semantic perturbations of textual~\cite{ruane2018botest,guichard2019assessing
} 
or voice conversations~\cite{iwama2019automated,guglielmi2024help}. Other approaches applied adversarial attacks to automatic speech recognition systems~\cite{
bilika2024hello}, whereas very limited research specifically addressed task-based chatbot security testing~\cite{bozic2020interrogating}.

Empirical evidence is small-scale, 
where the considered chatbots are typically few, ranging from just one~\cite{vasconcelos2017bottester,ruane2018botest,guichard2019assessing,bozic2019chatbot,bozic2019testing,bovzic2022ontology} to rarely up to ten~\cite{gomez2024mutation}, often selected from previous papers without implementing any quality and relevance selection criteria. This is mainly due to the lack of datasets in the area that hinders progress, complicating empirical studies and comparisons among techniques.

To the best of our knowledge, the only relevant dataset of task-based chatbots available so far is the one used in the experiments with ASYMOB~\cite{canizares2024measuring,lopez2022asymob}, a framework built to evaluate and classify chatbots using conversational quality metrics, dating back to 2022
~\cite{asymob}. 
The chatbots in the dataset have not been selected based on their complexity, nor have they been analyzed, and many of them are outdated nowadays. 


This paper addresses this gap by presenting the \bigDatasetFullName (\bigDataset) dataset, a snapshot of the Rasa chatbots available on GitHub. \bigDataset derives from the analysis of \initialNumber repositories hosted on GitHub projects and represents the state of the practice in open-source task-based chatbot development with Rasa. Further, this paper presents the \datasetFullName (\dataset) dataset, a carefully curated dataset of task-based chatbots, which derives from \bigDataset, finally delivering a set of \BRASATONumber relevant chatbots selected according to criteria about \textit{dialogue complexity},     \textit{functional complexity}, and \textit{utility}. 

We focus on the Rasa platform since it is one of the most popular open-source platforms~\cite{abdellatif2021comparison,cao2023characterizing},
extensively used in industrial and commercial applications. The Rasa open-source repository on GitHub\footnote{\url{https://github.com/RasaHQ/rasa}} counts more than 20,000 stars, 750 contributors, and about 50 million downloads\footnote{\url{https://techcrunch.com/2024/02/14/rasa-an-enterprise-focused-dev-platform-for-conversational-genai-raises-30m/}}, with customers including Orange, T-Mobile, and N26\footnote{\url{https://rasa.com/customers/}}. 
The choice of targeting an open-source platform gives maximum flexibility for future studies, allowing research not only at the level of chatbots, but also on the interplay between chatbots and the hosting platform, and the challenges related to this interaction~\cite{cao2023characterizing}.

The selection and analysis of chatbots in \dataset is relevant to software reliability in many ways. All the selected chatbots extract and process inputs from user sentences. This is a key feature for research on security, functional testing, and robustness, which studies how chatbots respond to different classes of inputs. Further, we analyze the chatbots in \dataset to report the external services used. This aspect is again important for research on testing, security, performance, failure detection and propagation, and authorization and authentication, which can be properly assessed only with subjects of proper functional and architectural complexity. 
The experimental material, including the datasets and the tools for replicating and updating chatbot selection in the future, is available online~\cite{dataset}.



In a nutshell, this 
paper has the central goal of supporting empirical research on chatbot quality by delivering a large-scale, curated dataset of real-world Rasa chatbots, facilitating reproducible and meaningful studies. More specifically, the paper
provides four 
key contributions: 

\emph{i) The retrieval and analysis of the full spectrum of Rasa chatbots available on GitHub}: We indicate this set of chatbots as \bigDataset. \bigDataset consists of \GRCSNumber Rasa chatbots extracted on January 14, 2025. \bigDataset captures the state of the practice in terms of task-based chatbot development with Rasa, and its analysis provides a clear picture of the activity of the open-source community.

\emph{ii) The selection and analysis of the curated \dataset dataset}: We select a core of \BRASATONumber Rasa chatbots that are significant for dialogue complexity, functional complexity, and utility, that the software reliability research community can exploit. We analyze \dataset chatbots, providing descriptive data essential to support research, such as the number of intents and entities recognized by each chatbot (dialogue complexity), the number and type of third-party services used by chatbots (functional complexity), and the Rasa version used and the popularity of the hosting repository (utility).


\emph{iii) A tool for chatbot selection and analysis, and artifacts to reproduce the study}: To facilitate future research in the area and enable the continuous update of the datasets, we not only release publicly the \bigDataset and \dataset datasets, but we also share the toolchain used to select and analyze the chatbots, as well as the sheets with the intermediate results, allowing our work to be exploited by others (e.g., by modifying our tools) even for purposes that we cannot anticipate.  

\emph{iv) A small-scope experiment on reliability to demonstrate the practical utility and relevance of \dataset}: We selected ten chatbots from \dataset to collect preliminary evidence on their utility to the software reliability community.




The paper is organized as follows.
Section~\ref{sec:rasa} introduces both task-based chatbots and the Rasa architecture. Section~\ref{sec:methodology} presents the methodology adopted to collect and analyze the chatbots and to construct the datasets. Section~\ref{sec:dataset} presents descriptive statistics about the two datasets and the preliminary experiment with \dataset. Finally, Section~\ref{sec:related} presents the related work, followed by Section~\ref{sec:conclusions} for the conclusions. 

%% file: rasa.tex
Chatbots -- often referred to as virtual assistants or conversational agents -- are software applications that engage with human users via text, voice, images, or other forms of communication~\cite{adamopoulou2020chatbots}. 
They can be classified by their \textit{primary function} (i.e., informative, conversational, or task-based), by their \textit{domain scope} (i.e., general-purpose or domain-specific), and by the \textit{used technology} (e.g., natural language processors, speech recognition systems).
Task-based chatbots, in particular, are chatbots built to perform specific tasks through conversations, prioritizing the fulfillment of user intents, like booking an appointment or guiding a user through a process, rather than engaging them in open-ended dialogues~\cite{adamopoulou2020chatbots,grudin2019chatbots}.

Rasa~\cite{rasa} is one of the most popular open-source platforms for building multi-language task-based chatbots~\cite{abdellatif2021comparison,cao2023characterizing}. 
The platform, implemented in Python, offers full control over chatbot data and infrastructure, employs machine learning modules for natural language understanding, and natively supports integration with several messaging platforms (e.g., Telegram, Slack) and external services, including APIs for connections with existing testing tools, such as Botium~\cite{botium}. 

A Rasa chatbot is composed of the following main elements that encapsulate the concepts of a conversation~\cite{cao2023characterizing}, usually defined in a \emph{Domain} YAML file. 

\emph{Intents}: The objectives a user may have, which the chatbot interprets as tasks to accomplish; users communicate their intents through \emph{utterances}. For instance, the intent to \emph{order a pizza} can be expressed with the sentence (i.e., the utterance) ``\emph{I would like to order a pizza}".

\emph{Actions}: The actions a chatbot has to perform to accomplish a task. Actions can be 
responses/requests (e.g., asking ``\emph{What pizza would you like to order?}"), graphical elements to interact with, or calls to external services. 
We can distinguish between simple \emph{response actions}, which are plain text responses to textual or graphical inputs, and \emph{custom actions}, which are actions that run a functionality implemented in \emph{backend action files}.

\emph{Entities}: The data types that are introduced by a user in sentences and that are recognized and processed by the chatbot. For instance, the user's answer ``\emph{I want a Pepperoni pizza}" encapsulates the entity \emph{pizzaType} 
whose value is \emph{Pepperoni}. 

\emph{Slots}: Variables that are filled in with values during conversations (e.g., with the value of an entity) and that can be used to guide the flow of the conversation or trigger actions.

Furthermore, a Rasa chatbot is enhanced with \emph{Stories} (i.e., flexible examples) and \emph{Rules} (i.e., constrained paths) to define \emph{Flows}, which are the possible user-bot interactions composing conversations, described as sequences of intents and actions.

Rasa chatbots activate these components through the Natural Language Understanding (NLU) module and the Dialogue Manager (DM) module~\cite{rasa,cao2023characterizing
}.
The NLU module leverages pre-trained language models, tokenizers, and classifiers to extract intents and entities from user utterances. These are then passed to the DM module, which determines the next action to perform -- either by replying directly, by triggering an action that may involve external services, or by employing more sophisticated Natural Language Generation (NLG) modules. 


%% file: methodology.tex
This section presents the methodology used to collect the \bigDataset and the \dataset datasets. \bigDataset is a dataset of \GRCSNumber Rasa chatbots representative of the open-source chatbots available on GitHub on January 14, 2025. \dataset is a curated selection of \BRASATONumber relevant chatbots extracted from \bigDataset according to their dialogue complexity, functional complexity, and usability. 

\begin{figure*}[h]
    \centering
\includegraphics[width=0.9\textwidth]{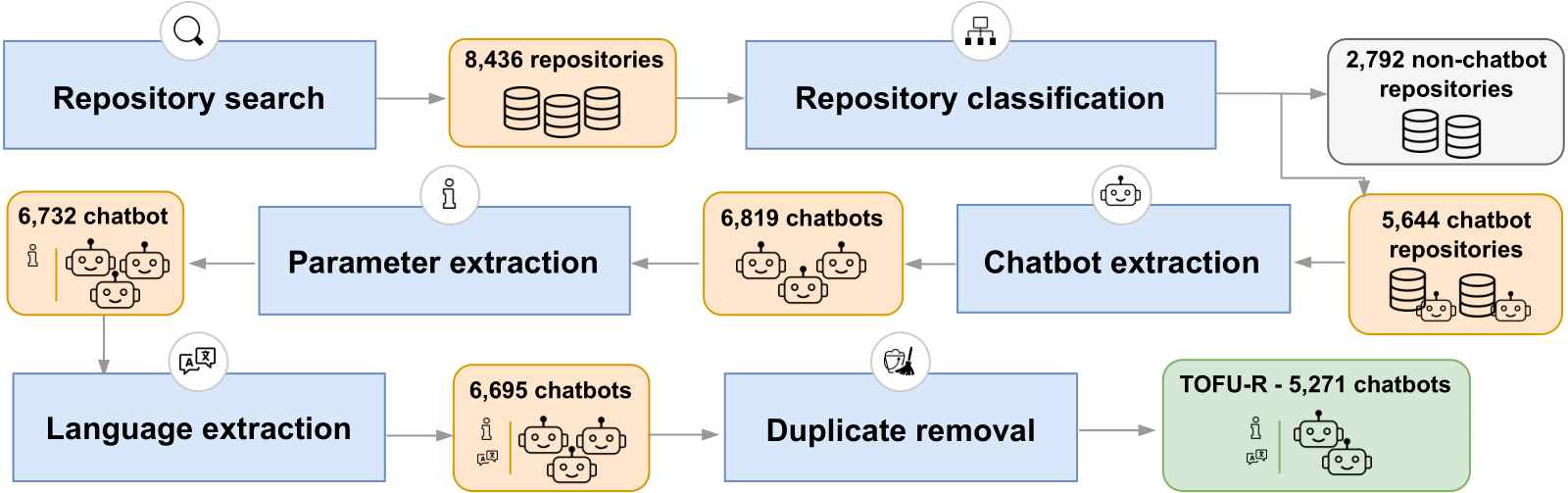}
    \caption{Methodology for the creation of the \bigDataset dataset.}
    \label{fig:dataset-methodology-wide}
    \vspace{-5mm}
\end{figure*}

\subsection{Creation of the \bigDataset dataset}
The construction of \bigDataset follows a six-step process, as illustrated in Figure~\ref{fig:dataset-methodology-wide}.

%

\subsubsection*{1) Repository search} In this step, we identified GitHub repositories that potentially include a Rasa chatbot. To this end, we used the GitHub REST repository API\footnote{\url{https://docs.github.com/en/rest/repos}} 
to search all publicly available repositories that on January 14, 2025 contained the keywords ``Rasa'' and ``chatbot'' in the title, README file, description, or topic list. We use this criterion since projects using Rasa typically mention the framework and its purpose (``chatbot'') in the repository metadata (e.g., title, README, or description). Moreover, these fields are indexed by the GitHub search engine, making it unlikely that any relevant repository is excluded.

This search resulted in \initialNumber repositories potentially hosting Rasa chatbots. 
To keep a reference to the repository version available on January 14, 2025, we retrieved the last commit's SHA identifier of each repository via GitHub REST commit API. This process collaterally identified 198 empty repositories, as they had no commits to retrieve. We thus ended up with 8,436 non-empty repositories.

\subsubsection*{2) Repository classification} The repositories retrieved in the first step may include actual Rasa chatbot projects, as well as repositories related to Rasa chatbots that do not contain a bot 
(e.g., Rasa frameworks extensions or chatbot development tools). 
Since the goal is to identify actual Rasa chatbots, we analyzed the retrieved repositories by checking for the presence of \textit{domain files}, which are required by actual chatbots to define their structural elements, such as intents and entities. 
Although chatbots typically include a single domain file named \textsc{domain.yml}, the name and number of these files may vary. To address this challenge, instead of searching for this specific filename, we searched for YAML files containing the \textsc{intents} field, which is a field required in any chatbot domain definition\footnote{\url{https://legacy-docs-oss.rasa.com/docs/rasa/domain/}}. Repositories with one or more of these files were classified as \emph{chatbot repositories}, while the rest were classified as \emph{non-chatbot repositories}. We downloaded the code of each bot to perform a local search.
 At the end of this step, we identified 5,644 chatbot repositories and 2,792 non-chatbot repositories, which were discarded.

\subsubsection*{3) Chatbot extraction} 
The objective of this step is to identify the individual chatbots stored in the selected repositories; in fact, the same repository may host one or more chatbots.
Of the 5,644 repositories resulting from the previous step, 943 contained more than one domain file, potentially hosting multiple chatbots. 
In particular, the presence of multiple domain files may correspond to various scenarios, such as
modularized domain definitions, coexistence of multiple chatbots within a single repository, presence of different chatbot versions in the same repository, or other ambiguous repository structures.  To capture these situations, we went through two main steps. 
%
%


\begin{itemize}[leftmargin=*]
\item \emph{removal of the invalid or misclassified domain files}: we automatically removed 1,494 domain files from 600 repositories, either because they were syntactically invalid or because they were other files incorrectly identified as domain files; 
60 repositories were finally discarded as they no longer contained valid domain files after this step.
\item \emph{identification of chatbots for repositories with multiple domain files}: 533 of the 5,584 remaining repositories had multiple domain files. 
Since distinct chatbots within a single repository are usually organized in different folders, we automatically identified the \textit{chatbot folders}, as the deepest-level folder in which the chatbot's domain files are saved. 
For instance, a repository with three domain files ``chatbots/a/domain1.yml",  ``chatbots/a/domain2.yml" and ``chatbots/b/domain1.yml" would be divided into two chatbots, ``chatbots/a" with two domain files and ``chatbots/b" with one domain file. Based on this analysis, 463 repositories with multiple domain files were split into 1,698 chatbots.
\end{itemize}



In total, 5,121 repositories could be directly mapped into individual chatbots as they had only one domain folder, summed up to the 1,698 chatbots derived from the multidomain repositories, lead to a total of 6,819 chatbots. 

\subsubsection*{4) Parameter extraction}
For each chatbot identified in the previous step, we analyzed each domain file individually to extract the key elements that characterize a chatbot's design and behavior. Specifically, we collected: the \emph{list of intents} recognized by the chatbot, the \emph{set of entities} that the chatbot can detect, the \emph{list of slots} and the \emph{type of variables} it can store, the \emph{set of actions} that the chatbot can execute, and the \emph{Rasa version} used, when available. After this analysis, 13 chatbots were discarded as their domain files were semantically incorrect (e.g., they assign values of incompatible type to some fields). We also sanity checked the chatbots with multiple domain files distinguishing three cases (111 chatbots had more than one domain file for an overall total of 328 domain files): (a) 8 chatbots had identical domain files, so we retained only one domain file for them; (b) 29 chatbots had domain modularized into multiple files with no overlap between their list of intents, entities, slots, and actions; for each of these chatbots, for analysis purposes, we merged all parameters extracted from each domain file to create a single composite domain; (c) the remaining 74 chatbots were discarded as their domain files were neither completely disjoint nor identical, making the identification of the overall domain unfeasible. This step resulted in 6,732 chatbots remaining.

\subsubsection*{5) Language extraction} To determine the language of each chatbot, we extracted two sets of languages: 
the \textit{training languages}, and the \textit{response languages}. 
The \textit{training languages} are the set of languages in which the training phrases are written.
To identify the training languages, we extracted the training phrases from two randomly selected non-default intents (the full set of training phrases could be too large to be processed) and used the Detect Language API\footnote{\url{https://detectlanguage.com/}} to identify the languages. 
The \textit{response languages} are the set of languages that the chatbot can use to answer questions.
All possible answers are defined in the response section of domain files. 
We applied the same process used to detect training languages to extract response languages. 
We identified 37 chatbots with neither training phrases nor responses, so we discarded them.

For the remaining 6,695 chatbots, we determined the overall languages based on the union of the training and response languages. 
To address potential inaccuracies in the language identification process (e.g., the API extracting both English and Italian from the sentence ``I would like a pizza"), we then manually checked the 837 chatbots with multiple languages and corrected 615 of them, leaving only 216 multi-language chatbots. Overall, we identified 59 different languages, with English being the most popular. We provide data about languages in Section~\ref{sec:dataset}.

\begin{figure*}[thb]
    \centering
    \includegraphics[width=0.9\textwidth]{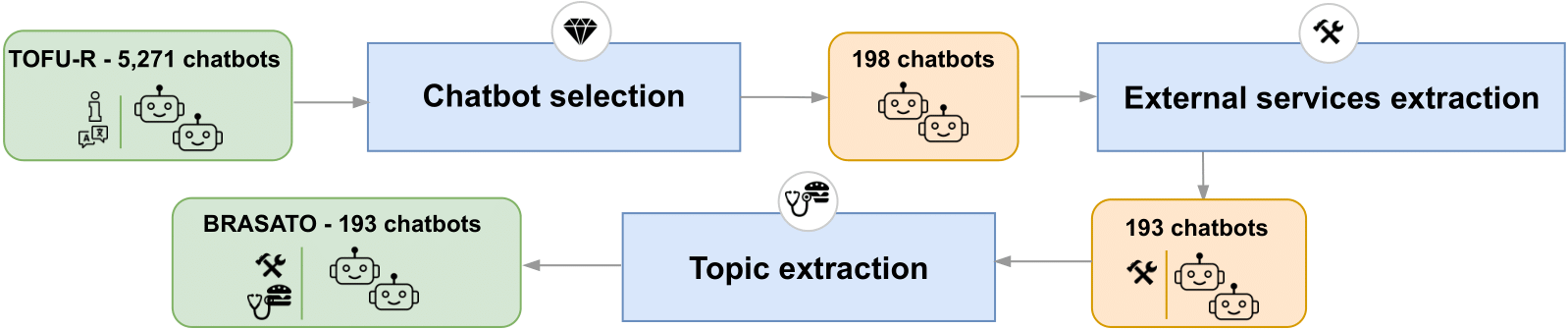}
    \caption{Methodology for the creation of the \dataset dataset.}
    \label{fig:brasato-methodology}
    \vspace{-5mm}
\end{figure*}

\subsubsection*{6) Duplicate removal} To ensure that \bigDataset is made of distinct chatbots, we finally checked and removed copies from the dataset. Two chatbots are copies of each other if they share the same list of conversational parameters extracted from the domains (i.e., intents, entities, slots, and actions) and the same configuration, training, and response languages. Since two chatbots with the same structure and language may differ for the implementation of the actions (e.g., the actions may have the same name in their domains but be implemented differently), we consider them as copies only if they have the same set of action files that share the same content for at least 95\% of the code identified with the Python library \textit{difflib}, which implements sequence comparison (we do not ask the code to be exactly the same since it might include irrelevant changes, such as whitespaces). 
For each group of copies, we selected the one to keep according to the following ordered list of criteria: Rasa version to favor up-to-date implementations, number of stars and forks to favor popular implementations, and creation date to select the original version of a chatbot. In this phase, we removed 1,424 copies, resulting in \GRCSNumber Rasa chatbots present in \bigDataset, finally representing the open-source chatbots currently available on GitHub~\cite{dataset}.

\subsection{Creation of the \dataset dataset}
While \bigDataset is representative of the activity in GitHub at large, researchers and practitioners are interested in working with relevant artifacts that can be used to meaningfully assess and compare software reliability approaches. To this end, 
we derived \dataset, a curated collection of \BRASATONumber chatbots selected from \bigDataset
according to criteria about the implemented dialogues, the available functionalities, and the utility of the chatbots. Figure~\ref{fig:brasato-methodology} presents an overview of the methodology used to define \dataset, which consists of three steps: \textit{chatbot selection}, to select the most relevant chatbots; \textit{external service extraction}, to associate each chatbot with its list of used external services; and \textit{topic extraction}, to also associate each chatbot with its application domain.

\subsubsection*{1) Chatbot selection}
We select the relevant chatbots in \bigDataset according to criteria about the \textit{dialogue}, the implemented \textit{functionalities}, and the \textit{utility} of the chatbot. 

The criterion about the dialogue guarantees that the selected chatbots can interpret a range of user requests, including the extraction of the entities (e.g., input values) that occur in the sentences.
This dimension relates to the set of intents, entities, and slots configured in the chatbots. 
To target chatbots that can handle significant dialogues, we selected those with \emph{at least one intent}, otherwise the chatbot would not be able to recognize any actual intention of the user, and \emph{at least one information retrieval component}, defined as either an entity or a slot, otherwise the chatbot would not process any user-generated input. 
In the \bigDataset dataset, 3,946 chatbots satisfy the selection criterion for the dialogue. 

We then consider the functionalities implemented by the chatbot. This dimension relates to the implemented actions, as they define the actual tasks a chatbot can perform, and the existence of a backend that implements functionalities that might be needed to execute the tasks. In fact, chatbots with dialogue capabilities that do not implement any action are chatbots that cannot deliver functionalities beyond the actual conversation.
To guarantee the selected chatbots implement and deliver some business logic, we select chatbots with \emph{at least one custom action implemented in a backend action file}. 
A total of 3,050 chatbots also satisfy this criterion.

Lastly, we narrowed our selection to chatbots that are appreciated and usable by a wide international audience. 

Since English is the most widespread and accessible language for the global community, we restricted the selection to the 2,186 chatbots that \emph{include English} in their configuration, training, and response languages (other languages can also be present). Since Rasa version 3.x is the standard from the end of 2021~\cite{cao2023characterizing}, we focused on up-to-date chatbots, \emph{removing all the chatbots implemented with older versions of Rasa},  narrowing the selection to 910 chatbots. As for community appreciation, the low popularity of a chatbot within GitHub may suggest that the chatbot is not interesting, valuable, or usable. We thus \emph{discarded all the chatbots with no stars}. This criterion led to the final selection of 198 chatbots.
\subsubsection*{2) External service extraction} The value and complexity of the functionalities provided by a chatbot largely depend on the services that can be executed as a consequence of a conversation.  Information about these services is particularly important for studies on software quality, reliability, and security that would have to consider them, for instance, when designing testing, dependability assessment, and security analysis approaches.  To better support these studies, we designed a procedure to identify the services and resources located outside the Rasa environment, which expand the boundaries of the system. Examples of external services are remote APIs (e.g., Wikipedia API, Gemini API), databases (e.g., MongoDB, MySQL), and non-Rasa files accessed during the execution; we do not consider the internal Python libraries in this list (e.g., pandas, matplotlib, keras).

\begin{figure}[t]
    \centering
    \includegraphics[width=\columnwidth]{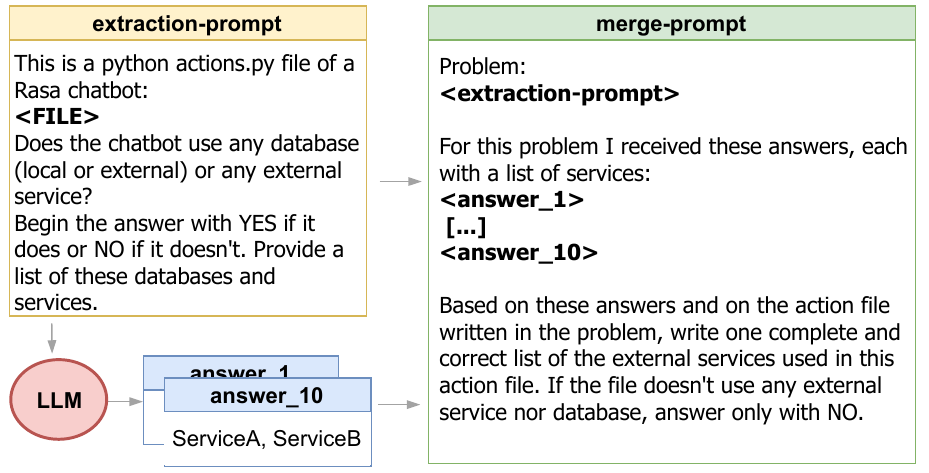}
    \caption{Extraction and Merge prompts.}
    \vspace{-6mm}
    \label{fig:prompts}
\end{figure}

Since retrieving these services while discriminating their nature and consistently assigning names to them would be quite hard and expensive with static analysis, we designed an LLM-based procedure that exploits both the analysis of the chatbot's code and the analysis of the README file in the repository to identify them.
For chatbots extracted from multi-chatbot repositories, we considered both the root README file and other README files found in the specific chatbot folders. To extract accurate and consistent information, we repeated each request for the identification of the used external services in the code or in a README 10 times, and then used an extra request to derive the final list from the analyzed element and the 10 responses. 
Figure~\ref{fig:prompts} shows the two prompt templates that we used. The extra request to merge results (merge-prompt) is important to finalize and harmonize the list, since the LLM can identify different services in each request and also refer to a same service with slightly different names that are semantically equivalent and that have to be normalized to avoid redundancies (e.g., SnowAPI and ServiceNow as two different ways of referring to a same service, where only one name should consistently occur in the list).

We used ChatGPT-4o to analyze our dataset. To identify the configuration of ChatGPT best suited for this task, we selected 11 chatbots with a variable number of external services (from 0 to 6) for which we manually determined the ground-truth list of services. We thus systematically submitted the queries to ChatGPT according to our prompt templates with multiple combinations of temperature\footnote{The temperature parameter controls the creativity of the model.} values (0, 0.25, 0.5, 1, 1.5, 2) and top-p\footnote{The top-p parameter influences the predictability of tokens in a sentence.} values (0.01, 0.15, 0.5, 1), for an overall total of 24 combinations. We measured the average precision, recall, and F-score over 5 repetitions for every configuration, and discovered that the optimal configuration, achieving 0.90 precision, 1.00 recall, and 0.93 F-score, is the one with temperature equal to 1 and top-p equal to 0.15. Interestingly, the few mistakes done by ChatGPT consist of incorrect and redundant services, but not in missing services. This suggests that a manual revision of the result produced by ChatGPT is likely to reveal all the services used by a chatbot.
We thus analyzed all 198 chatbots using this strategy, including the manual revision of the results, and identified 257 services from 405 action files and 174 services from 224 README files. 
With the manual check, we identified the discontinued services (e.g., api.covid19india.org), normalized service names to ensure consistent naming across chatbots, removed invalid services (e.g., internal method calls), and 
merged the sets of services extracted from the code and from the README files. After this check, we discarded 5 chatbots that depended on discontinued services and are not usable. We obtained a final list of 
330 external services distributed over 141 chatbots, with a maximum of 12 services per chatbot. We found that 52 chatbots do not use any external service. We kept them in the final selection, since chatbots can still deliver useful functionalities without relying on external services. 

\subsubsection*{3) Topic extraction} To accurately describe the chatbots selected in \dataset, we extracted their main topic (e.g., Finance, Medical, Sports) with ChatGPT. We designed a prompt with the name and description of the repository, the README files of the chatbot, if any, and the list of conversational parameters extracted from the domain (i.e., intents, entities, slots, and actions) and asked ChatGPT to return the topic of the chatbot. To minimize the variability in topic definition, we provided ChatGPT with the list of categories from Google Play Store\footnote{\url{https://www.searchapi.io/docs/parameters/google-play-store/categories}}, a widely used categorization of app domains, and asked to select a topic from the list or propose a new one if none of them was suitable. Given the simplicity of the task, this request was performed only once for each chatbot. We identified a total of 23 topics with this process, all selected from the given list. We manually inspected the results for a sample of 20\% of the topics to make sure ChatGPT returns accurate results.
The output of this step is \dataset, a curated collection of \BRASATONumber chatbots~\cite{dataset}.

%% file: dataset.tex
This section analytically describes the content of the two released datasets, discussing dialogue complexity, functional complexity, and utility of the chatbots. 

\subsection{Dialogue complexity}
\begin{figure}[t]
    \centering
    \includegraphics[width=\columnwidth]{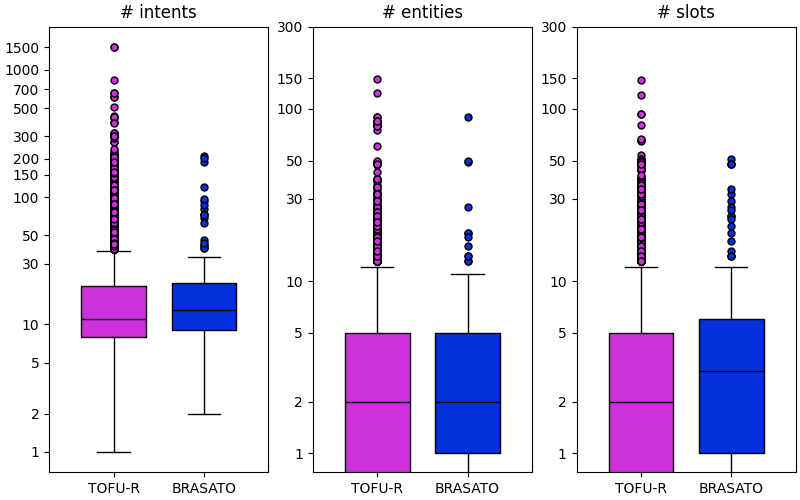}
    \caption{Dialogue parameters in \bigDataset and \dataset.}
    \vspace{-2mm}
    \label{fig:dialog-boxplots}
    \vspace{-3mm}
\end{figure}

Figure~\ref{fig:dialog-boxplots} shows the boxplots for the number of intents, entities, and slots in the chatbots available in the \bigDataset and in the \dataset datasets, using a logarithmic scale.

The number of intents in \bigDataset ranges from 1 to 1,515. The distribution is quite spread, with several single-intent chatbots and chatbots with a huge number of intents. 
The former are often toy examples, while the latter are typically Q\&A chatbots
providing fixed answers to predetermined questions, delivering no actual services beyond listing possible responses.

\dataset represents a much more curated selection than \bigDataset, without single-intent and large but irrelevant chatbots. The central part of the distribution is particularly interesting, since it includes chatbots that satisfy the selection criteria, still having between 9 (first quartile) and 21 (third quartile) intents, with a maximum of 210 intents. 

Similarly, \dataset includes fewer chatbots with a few or a huge number of entities or slots than \bigDataset. In fact, \dataset targets chatbots that actually process dialogues and data passed through dialogues, not including chatbots that do not process any data and do not deliver actual services. 

Across all the considered dimensions, the \dataset dataset has a higher median and narrower distribution than \bigDataset. 
Interesting cases of chatbots processing a significant number of intents, entities, and slots are the \textit{PI Rasa Chatbot}\footnote{\url{https://github.com/2nour/PI-Rasa-chatbot/tree/all\_langs}}, which uses 68 intents, 27 entities, and 21 slots to convert currencies and transfer money, or the \textit{Patient Appointment}\footnote{\url{https://github.com/VrajPatelK/Patient-Appointment}}, which supports appointment booking with 19 intents, 19 entities, and 15 slots. 

\vspace{-2.5mm}
\begin{figure}[t] 
\begin{minipage}[c]{0.55\columnwidth}
    \centering
    \includegraphics[width=\linewidth]{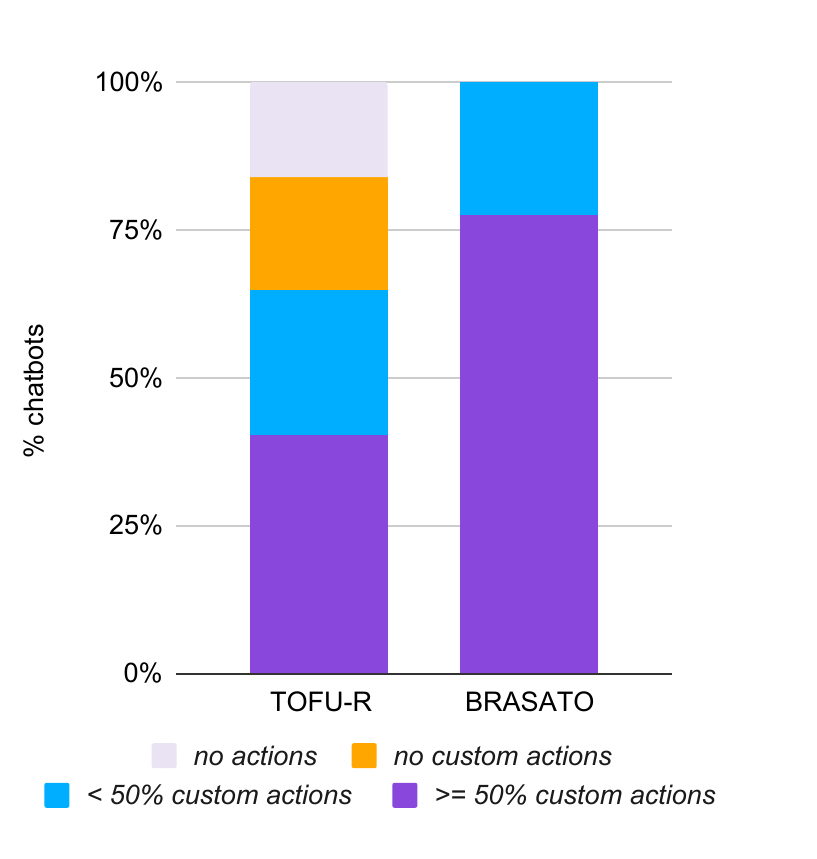}
    \vspace{-5mm}
    \captionof{figure}{Actions distribution in \bigDataset and \dataset.}
    \label{fig:actions-comparison}
\end{minipage}
\hfill
\begin{minipage}[c]{0.43\columnwidth}
    \centering
    \includegraphics[width=\linewidth]{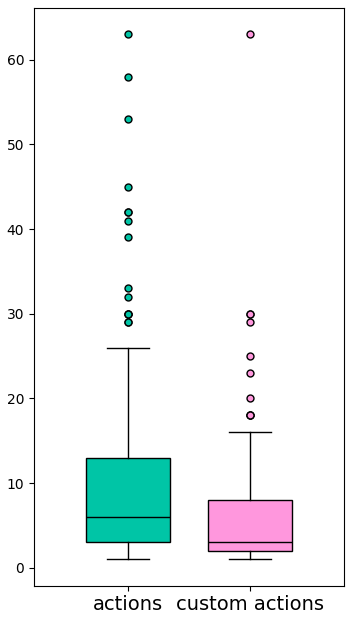}
    \vspace{-5mm}
    \captionof{figure}{Number of actions per chatbot in \dataset.}
    \label{fig:custom-actions-brasato}
\end{minipage}
\vspace{-5mm}
\end{figure}

\subsection{Functional complexity} To discuss the range of functions provided by the chatbots, we analyzed both the type of actions and the set of services provided by the chatbots in our datasets.

Figure~\ref{fig:actions-comparison} shows the density of chatbots with custom actions, that is, actions that run functionalities implemented in backend files. We could notice how \bigDataset includes many chatbots with no custom actions, and even no actions at all. This suggests that a large proportion of the chatbots available on GitHub are simple implementations of chatbots providing little or no functionality. On the contrary, all the chatbots in \dataset implement custom actions, with the vast majority mainly implementing custom actions. This confirms \dataset is the selection of the Rasa chatbots that provide actual functionalities from the whole set available on GitHub.

Figure \ref{fig:custom-actions-brasato} shows the number of actions and custom actions per chatbots. Indeed, all chatbots implement some actions, with several chatbots implementing a non-trivial number of them. Noticeably, more than 25\% of the chatbots implement at least 12 
actions and 8 
custom actions.

\begin{figure}[t]
    \centering
    \includegraphics[width=\columnwidth]{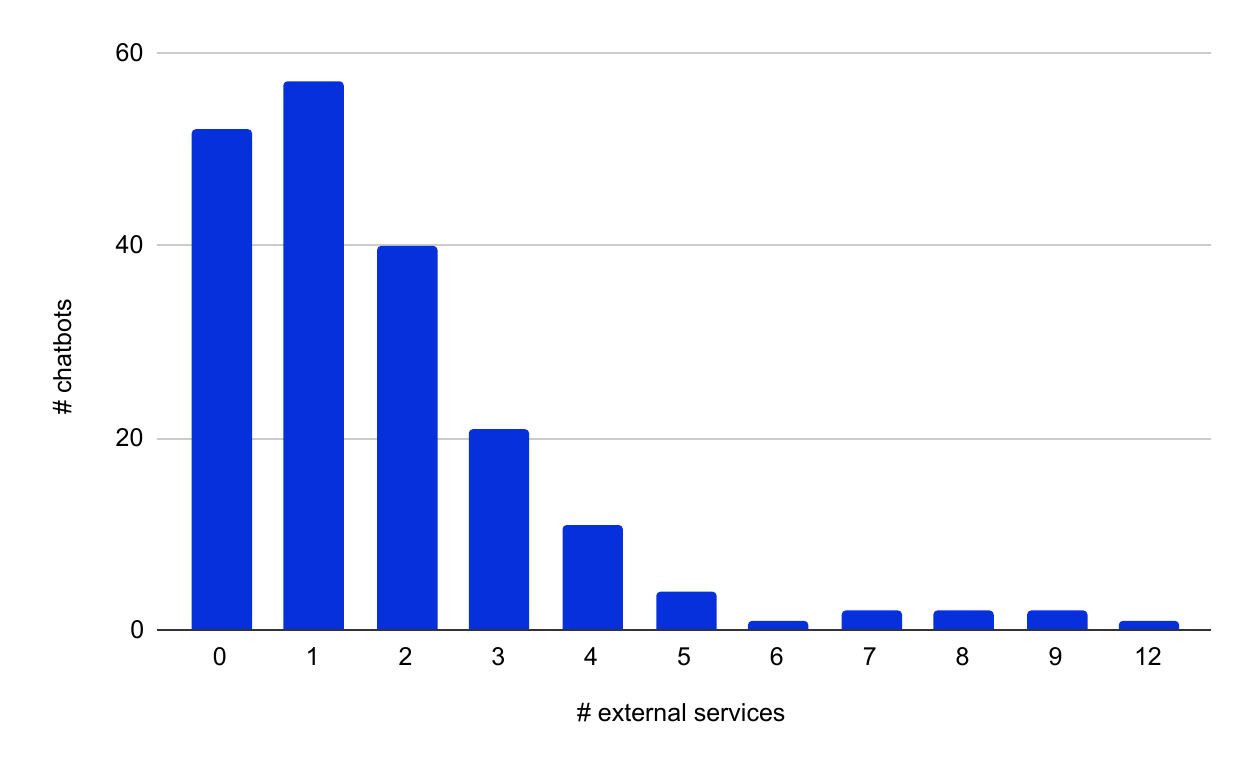}
    \vspace{-8mm}
    \caption{Number of chatbots by service number in \dataset.}
    \label{fig:services-grcs}
    \vspace{-4mm}
\end{figure}

We also analyzed the external services used by each chatbot. Figure~\ref{fig:services-grcs} shows the number of chatbots (y axis) that use a given number of services (x axis) in \dataset. The vast majority of the chatbots use external services (73\%), with a maximum of 12 services used. 

The services that appear with the highest frequency are: storage services (e.g., SQL databases, local files, and MongoDB are used in 19\%, 17\% and 5\% of the chatbots, respectively), OpenAI (10\% of the chatbots), OpenWeatherMap (6\% of the chatbots), and Gmail (5\% of the chatbots).

\subsection{Utility}
We discuss utility in terms of the language and Rasa version used by the chatbots, as well as the popularity of the repository that hosts the code of the chatbots.

\begin{table}[b]
\centering
\vspace{-4mm}
\caption{Number of chatbots by language count.} 
\vspace{-1mm}
\begin{minipage}[t]{0.48\linewidth}
\centering
\caption*{\bigDataset}
\vspace{-2mm}
\label{tab:languages-grcs}
\begin{tabularx}{0.9\linewidth}{c l}
\toprule
\textbf{\# languages} & \textbf{frequency} \\
\toprule
1 & 5060  \\
2 & 187   \\
3 & 21    \\
4 & 2     \\
7 & 1     \\
\bottomrule
\end{tabularx}
\end{minipage}
\hfill
\begin{minipage}[t]{0.48\linewidth}
\centering
\caption*{\dataset}
\vspace{-2mm}
\label{tab:languages-brasato}
\begin{tabularx}{0.9\linewidth}{c l}
\toprule
\textbf{\# languages} & \textbf{frequency} \\
\toprule
1 & 187  \\
2 & 4   \\
3 & 2    \\

\bottomrule
\end{tabularx}
\end{minipage}
\label{tab:multilanguage}
\vspace{-1mm}
\end{table}

Table~\ref{tab:multilanguage} shows the number of languages spoken by chatbots in \bigDataset and \dataset. We could notice as in GitHub there are nearly 4\% of chatbots that support multiple languages, with a maximum of seven languages supported by a single chatbot. We could thus claim that a strict minority of chatbots are designed to interpret multiple languages. 
A similar rate holds in \dataset, where about 3\% of the chatbots are multilingual (6 chatbots out of 193). Indeed, the current trend is designing monolingual chatbots.

\begin{figure}[t]
    \centering
    \includegraphics[width=\columnwidth]{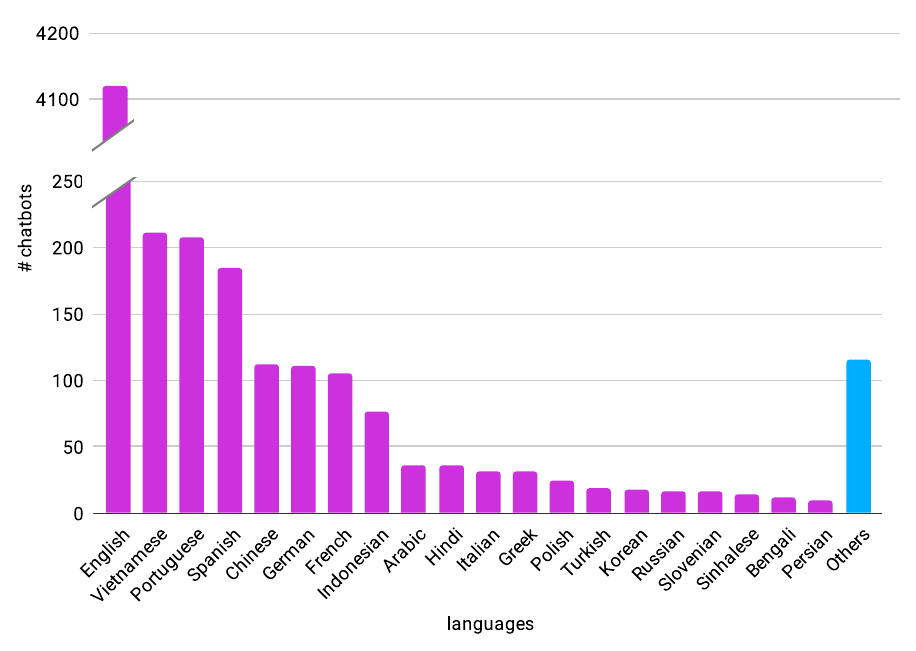}
    \vspace{-7mm}
    \caption{Language distribution in \bigDataset.}
    \label{fig:languages}
    \vspace{-5mm}
\end{figure}

Figure~\ref{fig:languages} shows the distribution of the chatbots by language for \bigDataset (multilingual chatbots are counted once for each language they support). By construction, all the chatbots in \dataset interpret English, in five cases in combination with other languages. 

Not surprisingly, by far, the most supported language is English (note that the plot is cut on the y axis, otherwise the disproportion of English chatbots would not make the rest of the plot visible). 
Although Vietnamese occurring as the second most frequent language may seem surprising, chatbots are extensively used in Vietnam for customer support and engagement, and their popularity is growing\footnote{\url{https://www.imarcgroup.com/vietnam-chatbot-market}}. The specific interest of developers for Rasa chatbots that speak Vietnamese is confirmed by the existence of
a Rasa forum for Vietnamese developers\footnote{\url{https://forum.rasa.com/t/rasa-community-in-vietnam/2993}} and by Google Trends reporting Vietnam as the $8^{\textit{th}}$ country in the world for searches for the keywords ``rasa" and ``chatbot" at the time of writing, since Rasa first release\footnote{\url{https://trends.google.it/trends/explore?date=2019-05-21\%202025-05-10\&q=rasa\%20chatbot\&hl=en}}. 

The fact that English is over-represented to the detriment of other languages may generate issues in the development of non-English chatbots and the availability of training data for relatively rare languages.

\smallskip

Table~\ref{tab:version-grcs} shows the distribution of chatbots in \bigDataset by Rasa version. The majority of chatbots support the most recent version, although a relevant number of chatbots are available for older versions of Rasa. 
Note that it was impossible to identify the precise Rasa version for 518 chatbots. Since 315 of these chatbots have not been updated after Rasa 3 was released, their version is surely older than Rasa 3.0 (Rasa $<$3.0 in Table~\ref{tab:version-grcs}).  The remaining 203 chatbots are also not likely to support Rasa 3.0 since they fail to satisfy the strong recommendation of explicitly specifying the supported version. However, to be safe, we classified them as undefined.

The gradually increasing number of chatbots available for Rasa 1, 2 and 3 reflects the increasing interest for the chatbot technology and on Rasa in particular.  
\dataset only includes Rasa 3 chatbots, which is the most recent and currently supported version of Rasa.

\begin{table}[]
\centering
\caption{Rasa version distribution in \bigDataset.}
\label{tab:version-grcs}
\begin{tabularx}{0.5\columnwidth}{l l l}
\toprule
\textbf{Rasa version} & \textbf{frequency} \\
\toprule
Rasa 3.x & 2218  \\
Rasa 2.x & 1386  \\
Rasa 1.x & 1149  \\
Rasa $<$3.0 & 315 \\
Undefined & 203  \\
\bottomrule
\end{tabularx}
\vspace{-1mm}
\end{table}

\smallskip

\begin{figure}[t]
    \centering
    \includegraphics[width=\columnwidth]{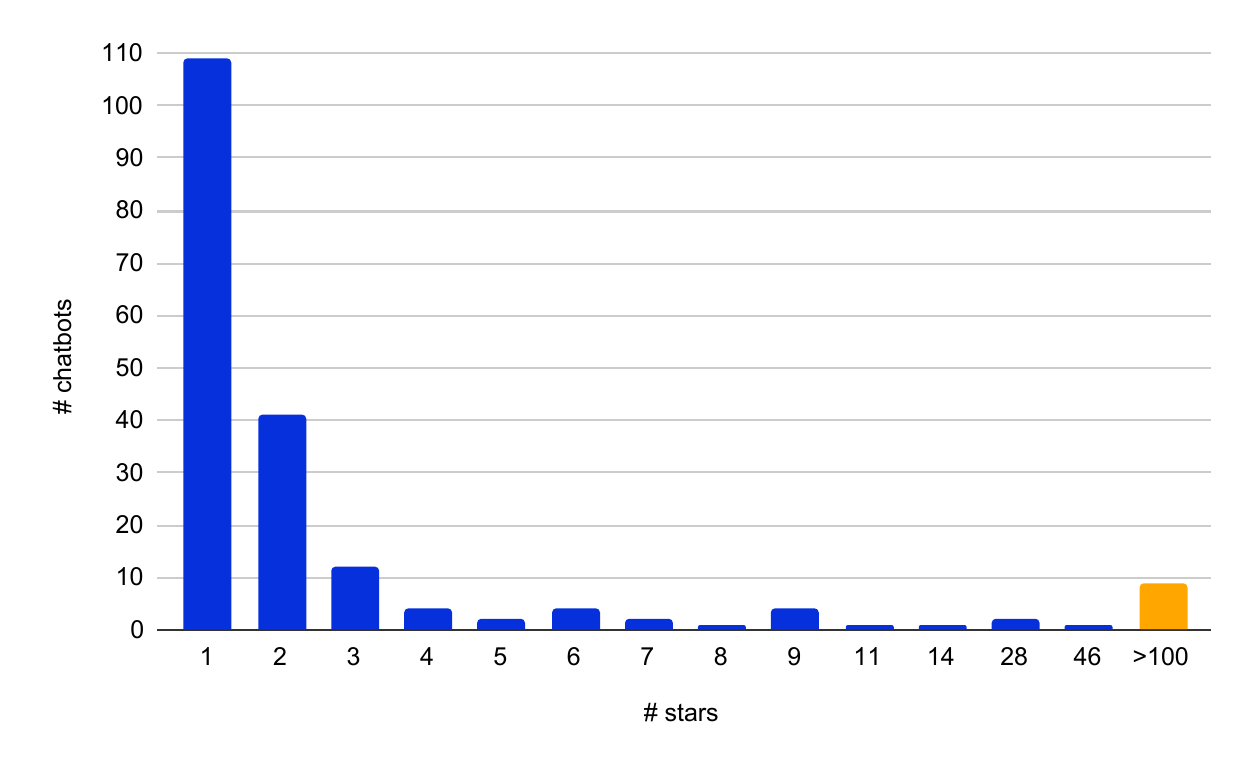}
    \vspace{-8mm}
    \caption{GitHub stars distribution in \dataset.}
    \label{fig:stars-brasato}
    \vspace{-4mm}
\end{figure}

Most Rasa chatbots available on GitHub are hosted in repositories without any stars, as only the 27\% of \bigDataset is starred. For \dataset, we only considered starred projects. Figure~\ref{fig:stars-brasato} shows the number of chatbots in \dataset hosted on repositories with a given number of stars. We could notice that although there are quite popular chatbots, such as Rasa official guideline chatbots (e.g., \textit{financial-demo}\footnote{\url{https://github.com/RasaHQ/financial-demo}} for finance services and banking, \textit{helpdesk-assistant}\footnote{\url{https://github.com/RasaHQ/helpdesk-assistant}} for IT help desk assistance), \textit{Expense Tracker Chatbot}\footnote{\url{https://github.com/Pallavi-Sinha-12/Expense-Tracker-Chatbot}} for expense tracking, and the Research and Development assistant \textit{Personal R\&D Lab}\footnote{\url{https://github.com/hongbo-miao/hongbomiao.com}}, several chatbots are yet relatively popular. However, chatbots hosted in repositories with few stars often implement interesting and useful services, such as \textit{Lesion Detection}\footnote{\url{https://github.com/shreyahunur/Lesion-Detection}}, a chatbot that assists gastroenterologists with polyp detection, \textit{Medical Chatbot}\footnote{\url{https://github.com/injusticescorpio/Medical-Chatbot}}, which supports hospital bookings and disease predictions, and \textit{F1 Stats Bot} that provides Formula 1 information from 1950 to the present day\footnote{\url{https://github.com/AlessandroManilii/F1\_Stats\_ChatBot}}.

\begin{figure}[t]
    \centering
    \includegraphics[width=\columnwidth]{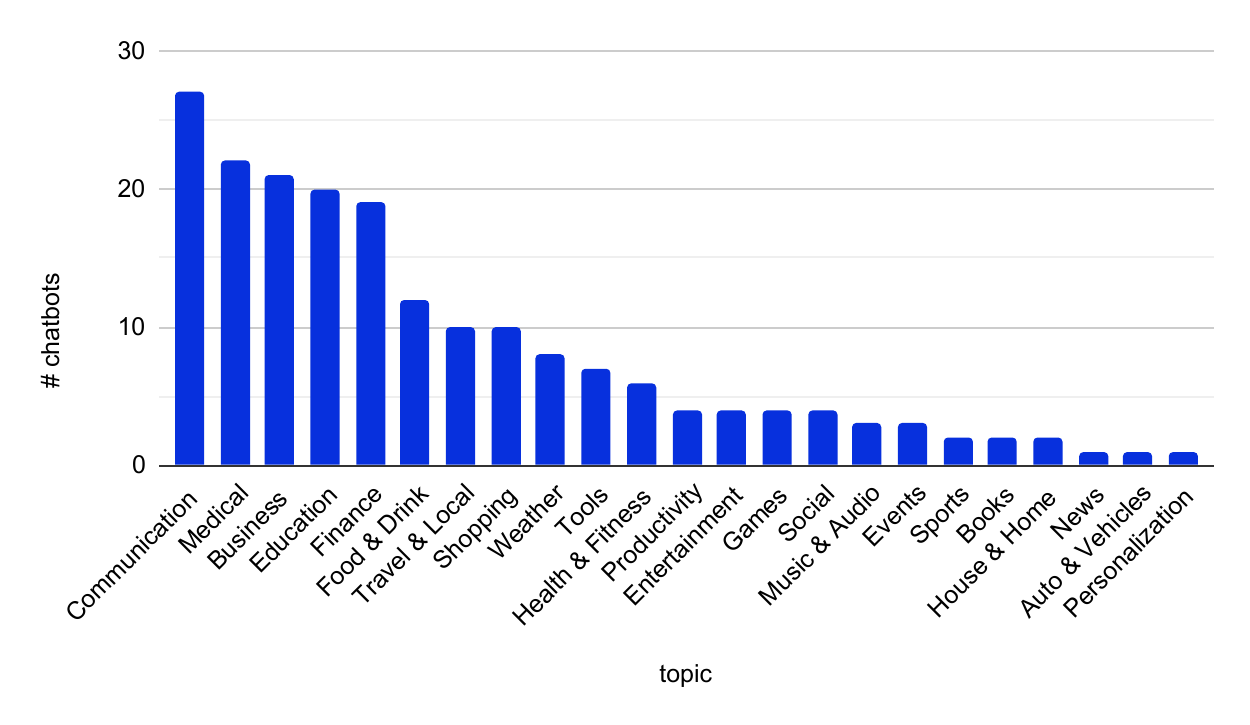}
    \vspace{-8mm}
    \caption{Topics distribution in \dataset.}
    \label{fig:topics-brasato}
    \vspace{-5mm}
\end{figure}

Finally, Figure \ref{fig:topics-brasato} shows the distribution of chatbot topics in \dataset, with \textit{Communication} (14\%), \textit{Medical} (11\%), \textit{Business} (11\%), and \textit{Education} (10\%) being the most common categories. 
Note that, although communication is a capability of all chatbots, Communication chatbots implement specific communication capabilities, such as supporting multiple languages, handling inventory or user data during conversations, and interacting with external APIs, such as OpenAI and Google APIs. It is also interesting to note that 18\% of the medical chatbots are Covid-19 related, since chatbots usage and popularity widely grew during the pandemic. 

\subsection{Preliminary Reliability Experiment on \dataset}

\dataset can support a broad range of use cases in the reliability domain. For instance, \dataset chatbots can be used to investigate the effectiveness of testing, security, fault tolerance, and robustness analysis techniques. 

To support these claims, we conducted a preliminary experiment on testing and security. 
We selected ten chatbots from the dataset\footnote{The selected chatbots are: \emph{AIDANXhang/CMPE\_252\_project\_AidBot}, \emph{Abishek-Suresh/Chatbot-Developer-Internship-Project}, \emph{DiorChoppa/financial-demo}, \emph{RasaHQ/helpdesk-assistant}, \emph{fritz007x/Multilingual-Chatbot-for-Customer-Service}, \emph{gowthamsenthil/Interview-schedule-bot}, \emph{hmalik123/SimpleRasaChatbot}, \emph{injusticescorpio/Medical-Chatbot\_\_Arista 2.O}, \emph{mammhoud/ChatHack}, and \emph{richkirk1/rasa-chatbot}.}, focusing on those using pre-built models for entity extraction (e.g., Duckling), relying on external services, and supporting categorical entities (i.e., entities that require a predefined set of possible values). 

In our testing experiment, we used the state-of-the-art test generation technique Botium~\cite{botium} to generate test cases for the selected chatbots, and
we observed (i) missing tests for intents with scarce or complex training phrases, or complex activation conditions, such as executing GUI actions (7 bots out of 10), (ii) poor coverage of entities handled by pre-built models or with categorical values not appearing in any training phrase (6 bots out of 10), and (iii) broken tests caused by special characters in annotated data, such as ``New\_Jersey'' that is truncated to ``New'' in a generated test (1 bot out of 10).

In our security experiment, we executed the Bandit\footnote{\url{https://github.com/PyCQA/bandit}} static analysis tool on the selected chatbots, to detect potential security vulnerabilities. We observed (i) unsanitized parameters extracted from conversations leading to possible SQL injections (3 bots out of 10), (ii) sensitive data exposure due to plain-text credentials (2 bots out of 10), and (iii) occurrences of the try-except-pass anti-pattern of catching exceptions without reporting the errors (1 bot out of 10).

These initial results show, on one hand, how tools must be further improved to thoroughly validate task-based chatbots, and, on the other hand, how \dataset can be used in reliability experiments. 

%% file: related.tex
Traditional quality assurance methods have proven challenging to apply to task-based chatbots, due to the difficulties in thoroughly exercising the conversational features and in formulating proper oracles to assess their correctness. So far, only a few methods have addressed functional or security assessment of task-based chatbots as a whole~\cite{cabot2021testing,lambiase2024motivations,li2022review}. 

Botium~\cite{botium} is a widely used automated quality assurance framework providing both automatic generation and execution of test cases for chatbots, supporting multiple chatbot design platforms, including Rasa~\cite{rasa}, Amazon Lex~\cite{amazon-lex}, and Dialogflow~\cite{dialogflow}. 
Some approaches studied the generation of tests that simulate user interactions~\cite{vasconcelos2017bottester,bozic2019chatbot}, while 
others defined coverage criteria about conversations to guide test case generation~\cite{rapisarda2025test,canizares2024coverage}.
A widely investigated approach consists of starting from a predefined set of conversations, for instance generated with Botium, and then applying metamorphic relations~\cite{bozic2019testing,bovzic2022ontology} or mutations to assess the robustness~\cite{bravo2020testing,guichard2019assessing} and security~\cite{bozic2020interrogating} of chatbots.
Mutation testing~\cite{gomez2024mutation,ferdinando2024mutabot} has also been explored in the context of chatbots.

Even though the aforementioned quality assessment works provide interesting contributions, their empirical assessment is often limited according to multiple dimensions:

\textit{Limited Number of Subjects}: the chatbots used in the experiments are usually fewer than five and always less than sixteen
, sometimes only relying on a single chatbot for the evaluation~\cite{bozic2019testing,bozic2019chatbot,guichard2019assessing,bovzic2022ontology,ferdinando2024mutabot,vasconcelos2017bottester,ruane2018botest,bravo2020testing};

\textit{Limited Subjects Relevance}: several chatbots present a simple structure with no actual actions and services, frequently consisting of just toy or demo applications 
that might miss important characteristics of realistic task-based chatbots~\cite{bozic2019testing,guichard2019assessing,ruane2018botest,vasconcelos2017bottester,bovzic2022ontology,bozic2020interrogating,bravo2020testing,rapisarda2025test,ferdinando2024mutabot,canizares2024coverage,gomez2024mutation};

\textit{Biased, Outdated, or Untraceable Sources}: the chatbots used in the experiments are created by the authors themselves, are no longer maintained or accessible, or have no specified origin~\cite{bozic2019chatbot,bozic2019testing,bovzic2022ontology,bozic2020interrogating,bravo2020testing,guichard2019assessing,ruane2018botest,vasconcelos2017bottester,gomez2024mutation,ferdinando2024mutabot,canizares2024coverage,rapisarda2025test};

\textit{In-house or Non-standard Technology}: the chatbots are developed using in-house or non-standard technology that can hardly be reused for follow-up studies~\cite{vasconcelos2017bottester,bozic2020interrogating}.

The \dataset dataset is designed to mitigate these issues and create a reference dataset that can be used to empirically study the effectiveness and ease the comparison of techniques designed for task-based chatbots.
By publicly sharing the tools we implemented to create the dataset and analyze the chatbots, it will be particularly easy to replicate the whole process or adapt the process to new requirements. 


The only related attempt to construct a dataset dates back to 2022, 
with the ASYMOB dataset~\cite{asymob,lopez2022asymob,canizares2024measuring}, consisting of a set of Rasa and Dialogflow chatbots. 
Unlike our work, the creation of the ASYMOB dataset does not employ any selection of chatbots based on dialogue complexity, functional complexity, and utility, only implementing the exclusion of malformed chatbots or chatbots not defined in English. 
In fact, only 40\% of the collected chatbots have at least one star or fork on GitHub. 
Information about the adopted languages, topics, conversational complexity, and external services is also not extracted, missing the opportunity to recommend subjects suited to specific experiments. 

Other repositories available on GitHub collect chatbots built for various purposes, such as providing templates\footnote{\url{https://github.com/park-sungmoo/rasa-chatbot-templates}} or supporting educational activities\footnote{\url{https://github.com/riteshkagale/Rasa-Chatbot-Intro}}. These datasets are often redundant, outdated, and lack a well-defined rationale for the selection of chatbots and the construction of the dataset. Conversational AI platforms, such as Dialogflow and Rasa, offer sets of pre-built agents, intended either as functional examples or as ready-to-use models to support chatbot development. Nevertheless, these agents typically fail to cover complete use cases.


In a nutshell, to the best of our knowledge, no other notable datasets consisting of curated full-fledged task-based chatbots that are also carefully analyzed from a dialogue, functional, and utility perspective have been publicly released.

%% file: conclusions.tex
Task-based chatbots are chatbots designed to deliver services and accomplish specific tasks using natural language. Despite their growing popularity, studies in the reliability domain have been mostly limited to a small number of subjects, sometimes of limited relevance.

This paper addresses this issue by presenting two datasets: \bigDataset, which is a snapshot of the current level of development of Rasa chatbots on GitHub, and \dataset, which is a curated dataset with significant chatbots that can be used to study and compare approaches. The two datasets are analyzed in terms of dialogue complexity, range of functionalities, and utility, providing an overview that researchers can exploit to make the best use possible of the dataset.

Future work concerns replicating the chatbot selection procedure targeting other platforms (e.g., Dialogflow), and performing periodic updates to keep the selection procedure and the datasets up to date. 
Finally, we plan to exploit the datasets to develop and assess quality control techniques.